# Unusual Sensitivity of Superconductivity to Strain in Iron-Based 122 Superconductors

J. S. Kim, G. N. Tam, and G. R. Stewart

Physics Department, University of Florida, Gainesville, FL 32611

**Abstract:** Co-doped $BaFe_2As_2$ has been previously shown to have an unusually significant improvement of $T_c$ (up to 2 K, or almost 10%) with annealing 1-2 weeks at 700 or 800 °C, where such annealing conditions are insufficient to allow significant atomic diffusion. While confirming similar behavior in optimally Co-doped $SrFe_2As_2$ samples, the influence on $T_c$ of strain induced by grinding to ~50 μ sized particles, followed by pressing the powder into a pellet using 10 kbar pressure, was found to increase the annealed transition width of 1.5 K by approximately a factor of ten. Also, the bulk discontinuity in the specific heat at $T_c$, $\Delta C$, on the same pellet sample was completely suppressed by grinding. This evidence for a strong sensitivity of superconductivity to strain was used to optimize single crystal growth of Co-doped $BaFe_2As_2$. This strong dependence (both positive via annealing and negative via grinding) of superconductivity on strain in these two iron based 122 structure superconductors is compared to the unconventional heavy Fermion superconductor $UPt_3$, where grinding is known to completely suppress superconductivity, and to recent reports of strong sensitivity of $T_c$ to damage induced by electron-irradiation-induced point defects in other 122 structure iron-based superconductors, $Ba(Fe_{0.76}Ru_{0.24})_2As_2$ and $Ba_{1-x}K_xFe_2As_2$. Both the electron irradiation and the introduction of strain by grinding are believed to only introduce non-magnetic defects, and argue for unconventional superconducting pairing.

I. Introduction

Since the discovery[1] of iron-based superconductivity (for reviews, see refs. 2-5), a number of unusual properties in these fascinating materials have been discovered. Sefat et al. discovered[6] superconductivity in $Ba(Fe_{1-x}Co_x)_2As_2$, with the peak of the superconducting transition temperature, $T_c$, dome vs composition at x=0.1 and the maximum $T_c$ equal to 22 K in as-prepared single crystals.

One of the properties of these materials which aroused interest was the substantial increase in $T_c$ in $Ba(Fe_{1-x}Co_x)_2As_2$ with annealing. Gofryk et al.[7], in the first annealing experiments, reported that crystals of $Ba(Fe_{1-x}Co_x)_2As_2$ attained $T_c$=25 K and a decrease in the transition width $\Delta T_c$ of 25% after 2 weeks at 800 °C. Kim et al.[8] further investigated $T_c$ vs annealing in $Ba(Fe_{1-x}Co_x)_2As_2$. They found in their optimized self-flux grown samples that as-prepared crystals with $T_c$=25 K reached $T_c$ values as high as 26.6 K via annealing for 1 week at 700 °C. With further optimization using finer gradations (in all, 17 different compositions between x=0.05 and 0.30) in Co concentration, Tam et al.[9] found as-prepared $Ba(Fe_{1-x}Co_x)_2As_2$ crystals with $T_c$=25.5 K, with comparable annealing[10] resulting in $T_c^{onset}$=27.2 K.

Such annealing, for 1 week at less[11] than 60% of $T_{melt}$ (i. e. without appreciable atomic diffusion) of pure $BaFe_2As_2$, resulting in such rapid increases in $T_c$ (~1.7 K or ~7% of $T_c$), seemed more effective than in other superconductors. (For example, annealing elements like Cd or Zn to narrow $\Delta T_c$ is done[12] at 95% of $T_{melt}$. The 14.4 K $T_c$ superconductor $YNi_2B_2C$, prepared by melting together the constituents, when annealed at 1200 °C for 5 days (72% of $T_{melt}$[13]), shows[14] no change either in $T_c$ or $\Delta T_c$.) The possibility that this unusual response of $T_c$ with annealing is a clue to the unusual superconductivity in iron-based superconductors motivated us in the present work to further investigate annealing in a second 122 structure system, $Sr(Fe_{1-x}Co_x)_2As_2$.

The results of this investigation as detailed below suggest a possible answer to the puzzle first posed by Gofryk et al. of why superconductivity in $Ba(Fe_{1-x}Co_x)_2As_2$ is improved so rapidly with relatively short annealing at only $\approx 60\%$ of $T_{melt}$.

## II. Experimental

Single crystal samples of nominal composition $Sr(Fe_{0.86}Co_{0.14})_2As_2$ (near optimal doping) were prepared using self-flux growth techniques as in refs. 8-9. The crystals nucleate out of self-flux (FeAs) during a slow cool (3 °C/hr) between 1200 and 900 °C, followed by a more rapid cooling (75 °C/hr) to room temperature. Crystals are then separated from the self-flux mechanically. A single crystal of mass 18.3 mg was chosen for a series of measurements on the <u>same</u> sample: measurement of magnetic susceptibility, $\chi$, and specific heat, C, on the unannealed crystal, annealing (700 °C for 2 weeks) of this crystal in an outgassed alumina crucible sealed via arc melting into a niobium cylinder containing an As vapor source[8-9], then measurement of $\chi$ and C on the annealed crystal.

Following the comparison of $\chi$ and C for the same single crystal, unannealed and annealed, the annealed crystal was ground in an inert atmosphere glove box in an agate mortar for 2-3 minutes for x-ray characterization. Before being x-rayed, the powder was pressed into a pellet (at 150,000 psi (10 kbar) to avoid poor thermal contact between the grains) and the susceptibility of the pressed powder pellet was measured. When these data showed severe degradation of the superconductivity, the specific heat on the pressed powder pellet was also measured. All of these measurements were on the same 18.3 mg single crystal, or the 12.9 mg pressed pellet from the powder therefrom.

The size of the powder making up the pellet (and of the powder made from a second crystal discussed below) was roughly determined by breaking it up as gently as possible with the blunt end of a wooden Q-tip and passing the powder through successively-sized sieves. Approximately half of the powder passed through a 270 mesh sieve (hole size 53 μ) and none of the powder passed through a 325 mesh sieve (hole size 45 μ).

In order to obtain an x-ray pattern on annealed powder to compare line widths and therefore strain between ground and annealed samples, a separate[15] crystal of mass 43 mg was ground in the inert atmosphere glove box. One part of the powder was x-rayed and measured via magnetic susceptibility and a second part annealed (without an As vapor source) for 2 weeks at 700 °C, and then measured by x-ray diffraction as well as by magnetic susceptibility. In addition, x-ray diffraction was measured on the unannealed single crystal to obtain the line widths of 00L reflections as discussed below.

III. Results

It was expected from the previous annealing work[8-9] on $Ba(Fe_{1-x}Co_x)_2As_2$ crystals that $T_c^{onset}$ and the bulk transition width of the specific heat of the single crystal of $Sr(Fe_{0.86}Co_{0.14})_2As_2$ would increase and narrow respectively upon annealing for 2 weeks at 700 °C. The susceptibility (Fig. 1) and specific heat (Fig. 2) data of the unannealed and annealed 18.3 mg single crystal of $Sr(Fe_{0.86}Co_{0.14})_2As_2$ confirm this expectation. $T_c^{onset}$ increases with annealing as measured by the susceptibility/bulk specific heat by ~1.5 K/0.9

K and the transition in the specific heat at $T_c$, $\Delta C$, sharpens considerably. These results are indeed comparable to those[7-9] in optimally doped $Ba(Fe_{1-x}Co_x)_2As_2$.

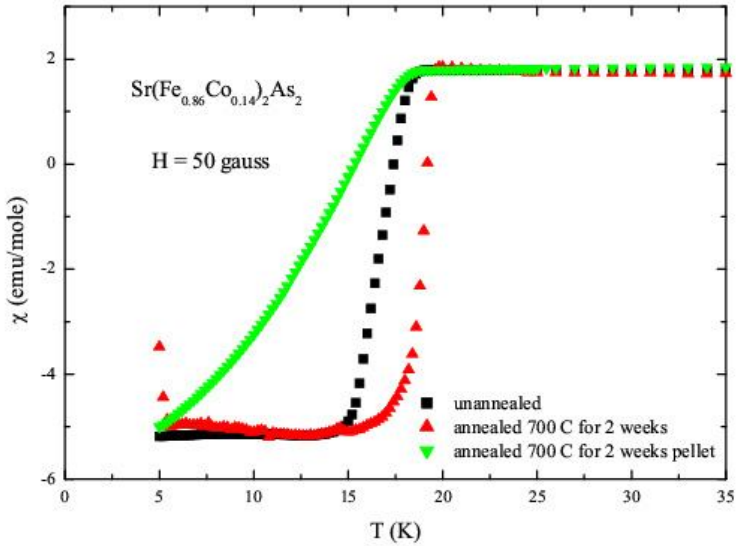

Fig. 1 (color online) Magnetic susceptibility, $\chi$, vs temperature of an 18.3 mg single crystal of $Sr(Fe_{0.86}Co_{0.14})_2As_2$ in three conditions: unannealed crystal (black squares), crystal after annealing at 700 ºC for 2 weeks (red triangles), and ground powder pressed into a pellet (green inverted triangles), >45 µ diameter, made from the annealed single crystal. $T_c^{onset}$ increases ~ 1.5 K with annealing, while the transition width, $\Delta T_c$, decreases from 2.4 to 1.5 K.

What was not known previously[7-9] is the very strong influence of grinding, followed by pressing into a pellet, on the superconductivity. As shown in Fig. 1, grinding[16] the annealed 18.3 mg crystal to a grain size of no smaller than 45 µ, followed by pressing into a pellet, results in a large increase (from ~1.5 K to over 14 K) in the transition width of the superconducting transition as measured by the magnetic susceptibility. Further, as shown in Fig. 3, specific heat of the pressed pellet of this same ground powder shows that the specific heat discontinuity at $T_c$, $\Delta C$, present in both the annealed and unannealed crystal (Fig. 2) is totally smeared out in the ground material.

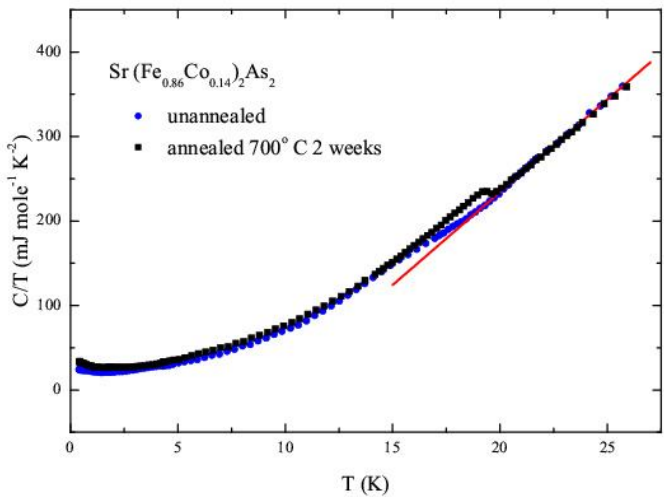

Fig. 2: (color online) Specific heat divided by temperature, C/T, vs temperature of an 18.3 mg single crystal of $Sr(Fe_{0.86}Co_{0.14})_2As_2$, unannealed (solid blue squares) and annealed at 700 ºC for 2 weeks (solid black circles.) $T_c^{onset}$ as measured by the bulk specific heat improves by ~ 0.9 K with annealing, comparable to work[8] on $Ba(Fe_{1-x}Co_x)_2As_2$. The red line is an extrapolation of the normal state data to below $T_c$. The finite intercept of C/T (T→0), defined as $\gamma_{residual}$, in these samples is ~20 mJ/molK$^2$, which is larger than that found[8] in Co-doped $BaFe_2As_2$ and could indicate the presence of some normal material in the crystal. However, the measured discontinuity in C at $T_c=19.5$ K, $\Delta C/T_c = 19$ mJ/molK$^2$ in the annealed sample is within 25% of that[8] for Co-doped BaFe$_2$As$_2$.

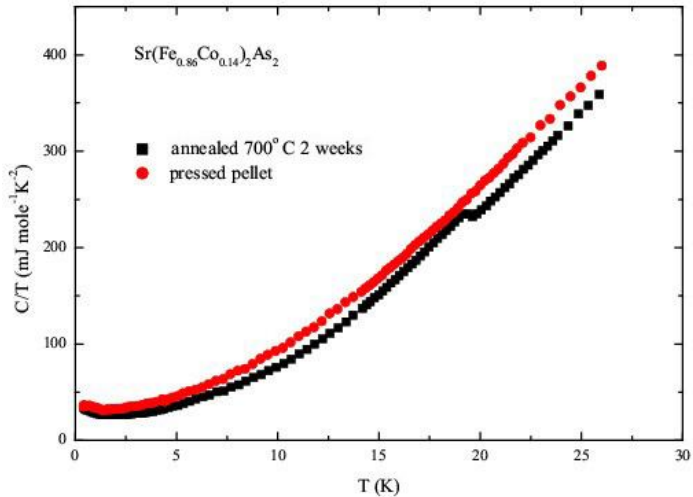

Fig. 3 (color online) Specific heat divided by temperature, C/T, versus temperature for annealed single crystal, 18.3 mg (solid black squares) and a pressed pellet of the ground powder, 12.9 mg (solid red circles), >45 μ diameter, from the same annealed crystal of Sr(Fe$_{0.86}$Co$_{0.14}$)$_2$As$_2$. The absolute error bar for the pressed pellet data above 20 K is ±5% vs ±3% for the single crystal due to the larger addenda contribution (38 % vs 17 %). However, the relative precision (± 1-2 %) between the two measurements is sufficient to state that the larger C/T for the pressed pellet sample is qualitatively correct.

In order to quantify the amount of strain introduced by the grinding, a separate 43 mg Co-doped SrFe$_2$As$_2$ crystal was ground[15] in the same fashion, and some of the (homogenized) powder was annealed for 2 weeks at 700 °C. We then measured x-ray diffraction and susceptibility on a portion[15] of the unannealed powder from this second crystal as well as on an annealed portion of the powder. The susceptibility of the unannealed starting crystal and on the ground powder are consistent with the results in

Fig. 1, while the susceptibility data of the annealed powder are consistent both in $T_c^{onset}$ and transition width with the annealed single crystal shown in Fig. 1, i. e. the annealed powder

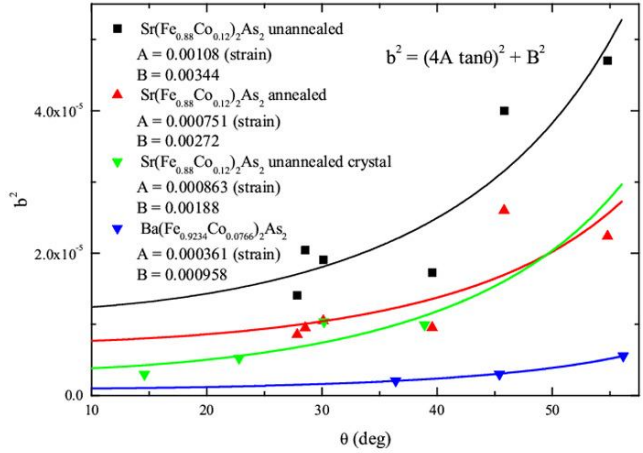

Fig. 4 (color online) Analysis of the line width, b (radians), vs angle of the xray reflections (hkl) to determine strain of unannealed (black squares) and annealed (red triangles) powder, as well as – using (00L) reflections - of an as-grown single crystal (green inverted triangles) of $Sr(Fe_{0.88}Co_{0.12})_2As_2$ and of a cooled-to-600 °C crystal (i. e. quasi-annealed) of $Ba(Fe_{0.9234}Co_{0.0766})_2As_2$, blue inverted triangles. For example, for the (0010) line the full width (when plotted vs Θ) at half maximum in units of $10^{-3}$ radians for the four different samples is 4.15, 3.09, 3.15, and 1.45 respectively. 'B' is the instrumentally caused line broadening. Errors bars for strain are ±0.00005 except for the Co-doped Ba 122 sample, where the error bar is only ±0.00002. (Note that a term ($\{0.9\lambda/Dcos\Theta\}^2$) in the equation for the line width that involves the particle size, D, is omitted since, with D~50 μ, the term is negligible.)

$T_c$ and $\Delta T_c$ are improved vis-à-vis the unannealed single crystal. Analysis[17] (Fig. 4) of the x-ray line widths of various (hkl) reflections of the annealed and unannealed powders between 55 and 110 degrees 2Θ results in a strain ε for the annealed powder of 0.0008 ± 0.0001 and for the unannealed powder of 0.0011 ± 0.0001, a small but - as evident from Fig. 4 – easily measurable[18] difference. Thus, although the annealing does cause a change in the amount of strain in the material, the small amount of the difference indicates a high sensitivity of the superconductivity to strain. Consistent with the susceptibility result (not shown) just discussed (that $T_c$(unannealed crystal) < $T_c$(annealed powder)), the analyzed strain in Fig. 4 in the unannealed single crystal of $Sr(Fe_{0.88}Co_{0.12})_2As_2$ is indeed slightly larger[19] than that of the annealed powder, 0.000863 vs 0.000751.

These results make it evident that the superconductivity in Co-doped $SrFe_2As_2$ is a.) very sensitive to strain and b.) the strain present in ground powder, with its very broadened $\Delta T_c$ and $\Delta C \to 0$, vs that in annealed material differs by a relatively small amount (~ 30%). Thus, the question arises: what is the minimal amount of annealing necessary to improve the superconductivity in unannealed single crystals? Phrased another way, what minimal further heat treatment on the as-grown self-flux crystals (with the slow (3 °C/hr) cooling halted at 900 °C) is necessary to remove the strain which the present work implies is introduced by cooling from 900 °C to room temperature at 75 °C/hr? Ref. 8 states that annealing at 600 °C has essentially no effect on $T_c^{onset}$ or $\Delta T_c$ in Co-doped $BaFe_2As_2$, therefore presumably removing this small amount of residual strain in the as-grown single crystals cooled slowly to 900 °C requires thermal treatment above 600 °C.

In order to make a first attempt at answering this question, and in a different 122 structure iron based superconductor in order to broaden the applicability of these results, we undertook the following. To verify indeed that the strain involved is produced by cooling at 75 °C/hr the as-grown crystals from 900 °C to somewhere above 600 °C (based on the ref. 8 result), we have reproduced/altered the growth procedure in our previous thorough study of annealing in Co-doped $BaFe_2As_2$ (refs. 8 and 9) for x=0.0766 (a composition slightly below that of optimal doping) as follows. Two batches of $Ba(Fe_{0.9234}Co_{0.0766})_2As_2$ crystals were grown in self flux, one heated to 1200 °C, cooled at 3 °C/hr until 900 °C, followed by cooling at 75 °C/hr to room temperature (the original procedure, followed also herein for Co-doped $SrFe_2As_2$.) The second batch was identical in every respect except it was cooled from 1200 °C down to 600 °C at 3 °C/hr, and then at 75

°C/hr to room temperature. This extra temperature region of slow cooling did not result in larger crystals, since the crystals have already formed[20] by 900 °C, but it adds slow cooling (roughly equivalent to annealing for the same length of time at a fixed intermediate temperature) over a period of about 3 days from 900 down to 700 °C. The susceptibility of single crystals from both batches is shown in Fig. 5. Clearly, the strain removed by annealing at 700 °C for 2 weeks in the present work, or at 700 °C for 1 week as in refs. 8-9, can also be removed by merely cooling at 3 °C/hr further down in temperature, past the previous 900 °C changeover-in-cooling rate point, to 600 °C. In fact, as shown in the strain analysis graph, Fig. 4, the cooled-to-600 °C crystal of $Ba(Fe_{0.9234}Co_{0.0766})_2As_2$ shows a strain only half of that of the annealed for 2 weeks at 700 °C powdered sample – arguing for the effectiveness of the slow cooling procedure.

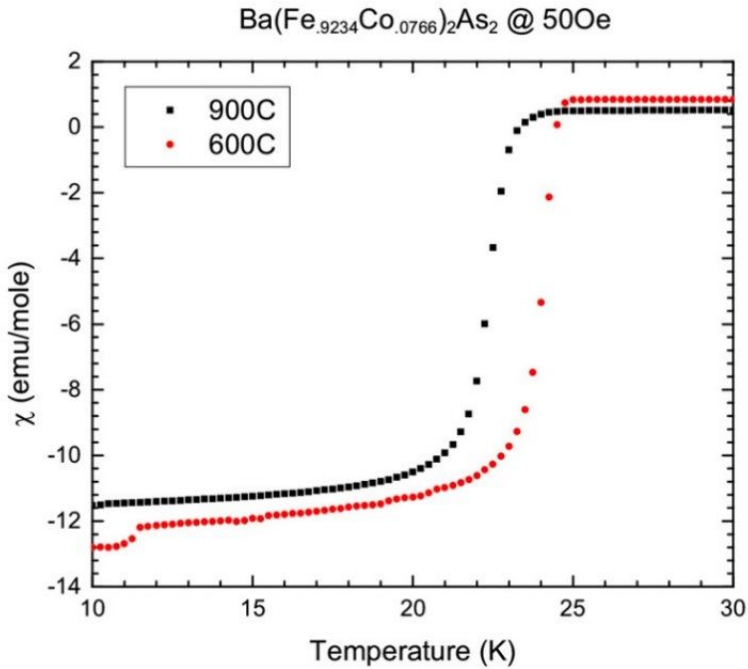

Fig. 5 (color online) Magnetic susceptibility, $\chi$, vs temperature for single crystals with the nominal composition of $Ba(Fe_{0.9234}Co_{0.0766})_2As_2$ prepared either by cooling at 3 °C/hr from 1200 °C to 900 °C, followed by cooling at 75 °C/hr to room temperature (black points) or by cooling at 3 °C/hr down to 600 °C, followed by cooling at 75 °C/hr to room temperature (red points). The difference in $T_c$ (either midpoint or onset) is approximately 1.7 K, the same as found[9] after 1 week at 700 °C annealing as-grown crystals slow cooled down to 900 °C. Note the somewhat sharper onset to superconductivity upon cooling in the red points, whereas the black data is somewhat more rounded.

**IV. Discussion and Conclusions**

The first conclusion that can be reached is the solution to the puzzle raised by the previous annealing work[7-9] on Co-doped $BaFe_2As_2$: why does annealing 1-2 weeks at 700 °C, only 60% of $T_{melt}$ have such an important effect on $T_c^{onset}$ and the bulk specific heat transition width, $\Delta T_c$? Clearly, the superconductivity in both the Co-doped $SrFe_2As_2$ and $BaFe_2As_2$ 122 iron based superconductors - and presumably other 122's and 111's as well (although see the discussion below of $CaFe_2As_2$) - is extremely sensitive to strain as shown by the results presented above.

Before we discuss why $T_c$ and $\Delta T_c$ are in Co-doped $SrFe_2As_2$ and $BaFe_2As_2$ are so sensitive to strain, we now make a digression in order to discuss whether the current results will apply to Co-doped $CaFe_2As_2$. First, it is important to note that results exist detailing how annealing affects the <u>normal state</u> properties in all three 122's. Annealing (at either 350 or 700 °C between 1 and 30 days) has been shown[21] to have a relatively small (1-6 K) effect on the magnetic and structural transitions for $BaFe_2As_2$ and $SrFe_2As_2$ (at 135 K and 200 K respectively), and to change the c-axis lattice parameters at room temperature by less than 0.01 Å. This is in stark contrast to $CaFe_2As_2$ (quenched from 960 °C in order to decant the crystals from the FeAs self-flux), where annealing at 400 °C for 1 week changes[22] the low temperature structure from a non-collapsed tetragonal phase below 100 K to an orthorhombic, antiferromagnetic state below 170 K, and increases[21] the c-axis lattice parameter by 0.152 Å. As an explanation for the 'extreme case' of $CaFe_2As_2$, ref. 22 explains that the effect of annealing is to remove very fine, ~10 nm, precipitates whose average strain field mimics the effect of 0.4 GPa pressure, and to allow the formation of the necessary-for-superconductivity orthorhombic antiferromagnetic state. Since a.) refs. 8 and 9 found that annealing at temperature *above* 600 °C is necessary to improve $T_c$ in

Ba(Fe$_{1-x}$Co$_x$)$_2$As$_2$ (where the current work tells us that the annealing is causing the removal of strain harmful for superconductivity), and since b.) ref. 23 finds that annealing at 600 °C and above in Co-doped CaFe$_2$As$_2$ forms the non-collapsed tetragonal, non-magnetic, inimical-to-superconductivity phase, the removal of strain from grinding ⇒ improved T$_c$ results presented here in Sr(Fe$_{1-x}$Co$_x$)$_2$As$_2$ likely cannot be used to optimize superconductivity (T$_c$~16 K) in Co-doped CaFe$_2$As$_2$. It would be interesting to measure the amount of strain present in the rather low T$_c$ Co-doped CaFe$_2$As$_2$ samples – is the strain larger than seen in the higher T$_c$ Sr(Fe$_{1-x}$Co$_x$)$_2$As$_2$ and Ba(Fe$_{1-x}$Co$_x$)$_2$As$_2$, or is the explanation the difference[23] in the order of the magnetic behavior in CaFe$_2$As$_2$ (strong first order) and the lack[23] of coexistence between magnetism and superconductivity anywhere in Co-doped CaFe$_2$As$_2$?

Returning now to our main question: why are T$_c$ and ΔT$_c$ in Co-doped SrFe$_2$As$_2$ and BaFe$_2$As$_2$ so sensitive to strain? Such a strong dependence of superconductivity on strain in Co-doped Ba and Sr 122 is indicative of an unconventional superconducting mechanism. One well known example[24] of a very strain sensitive superconductor is UPt$_3$, where grinding[24-25] totally destroys superconductivity (from a T$_c$ of ~ 0.5 K to below 0.05 K as measured by susceptibility.) The f-wave pairing symmetry in UPt$_3$ is expected to be very sensitive to damage and defects.[26]

Theory[5] suggests that the pairing mechanism, the so called s± scheme where the order parameter changes sign between different sheets of the Fermi surface, favored for the iron-based superconductors is also extremely sensitive to defects, not just magnetic defects as are known to degrade conventional superconductors but also including non-magnetic[27] defects introduced by grinding. Thus, the original intent of the present work – to see if understanding the unusually rapid improvement of $T_c^{onset}$ with annealing at only 60% of $T_{melt}$ for just 1 week in $BaFe_{2-x}Co_xAs_2$ could shed light on the superconductivity in iron based superconductors – has produced evidence for extreme sensitivity of the superconductivity to defects introduced via grinding. This is reminiscent of the behavior of the known unconventional superconductor $UPt_3$.

Another way to introduce non-magnetic defects in the lattice is via electron irradiation, which has been performed on 122 iron superconductors, $UPt_3$, as well as in the unconventional cuprate high temperature superconductors – thus allowing a quantitative intercomparison among all three. Electron irradiation, using 2.5 MeV electrons, by approximately $1.1 \; 10^{19}$ e/cm$^2$ gives $T_c$ reductions from the unirradiated $T_{c0}$ ($T_c/T_{c0}$) of 0.84 for[28] $UPt_3$, 0.80 for[29] $Ba(Fe_{0.76}Ru_{0.24})_2As_2$, 0.87/0.66 for[30] $Ba_{1-x}K_xFe_2As_2$ (x=0.19/0.34) and 0.92 for[31] $YBa_2Cu_3O_7$. These results are consistent with the extreme sensitivity of the superconductivity in Co-doped $SrFe_2As_2$ to grinding and the induced strain therefrom found in the present work. As well, the relief of a small amount of strain by replacing a 75 ºC/hr cooling from 900 to 600 ºC with 3 ºC/hr cooling and the concomitant increase of $T_c^{onset}$ by ~ 1.7 K in Co-doped $BaFe_2As_2$ is further consistent with the electron irradiation evidence[29-30] for the extreme sensitivity of iron based superconductivity, with presumed[29-30] s± pairing symmetry, to non-magnetic defects.

Acknowledgements: Work at Florida performed under the auspices of the US Department of Energy, Basic Energy Sciences, contract no. DE-FG02-86ER45268. Helpful conversations with P. Hirschfeld are gratefully acknowledged.

unavoidable contamination by chemical reaction. The nominal composition of the sample used for x-rays was $Sr(Fe_{0.88}Co_{0.12})_2As_2$. The identical growth procedure as for $Sr(Fe_{0.86}Co_{0.14})_2As_2$ produced larger crystals (with the same $T_c$ and $\Delta T_c$ since both compositions are at the peak of the superconducting dome) for this composition for grinding into sufficient amounts of powder.

16. Since the 122 iron based superconductors such as Co-doped $BaFe_2As_2$ and Co-doped $SrFe_2As_2$ are quite brittle, such grinding involves qualitatively (this will be quantified via x-ray line width analysis in later discussion) only a small application of force.

17. B. D. Cullity, Elements of X-Ray Diffraction, (Addison-Wesley, Reading, Mass., 1978); http://www.h-and-m-analytical.com/pdfs/size_strain.pdf

18. The differences in the high angle x-ray line widths in the ground and annealed powders of $Sr(Fe_{1-x}Co_x)_2As_2$ vary from 0.0016 radians (20% of the unannealed line width) at 55 degrees $2\Theta$ to 0.004 radians (30% of the unannealed line width) around 109 degrees $2\Theta$, i. e. these differences are easily measurable.

19. These two rather close values of strain each have error bars that make them almost equivalent. Unfortunately, the single crystal (00L) reflections, except for (008) are relatively faint and require long counting times for precision.

20. In A. F. May, M. A. McGuire, J. E. Mitchell, A. S. Sefat, and B. C. Sales, Phys. Rev. B **88**, 064502 (2013), Co-doped $BaFe_2As_2$ self-flux grown crystals are slow cooled (2 ºC/hr) only down to 1090 ºC, and then cooled more rapidly than 75 ºC/hr to room temperature.

21. B. Saparov and A. S. Sefat, Dalton Trans. 43, 14971 (2014).

which is consistent with increased magnetic behavior upon grinding.  However, such behavior does not appear to be affecting $T_c$ or $\Delta T_c$.